\documentclass{elsart}
\usepackage{amsfonts}
\usepackage{latexsym}

\newcommand{\emet}{{\em et al.}}
\newcommand{\be}{\begin{equation}}
\newcommand{\ee}{\end{equation}}
\newcommand{\bea}{\begin{eqnarray}}
\newcommand{\eea}{\end{eqnarray}}

\newcommand{\EA}{{\textit{et al.}}}
\newcommand{\IE}{{\textrm{i.e.}}}
\date{\today}
\begin{document}
\hfill{\small NT@UW-06-01}

\hfill{\small FZJ--IKP(TH)--2006--2}

\begin{frontmatter}

\title{
 Realistic Few-Body Physics
in  the $dd\rightarrow \alpha\pi^0$ Reaction}

\author{A.~Nogga$^1$, A.~C.~Fonseca$^2$, A.~G{\aa}rdestig$^3$, C.~Hanhart$^1$, 
            C.~J.~Horowitz$^4$,} 
\author{G.~A.~Miller$^5$, J.~A.~Niskanen$^6$, and U.~van~Kolck$^7$}

{\small 
$^1$ Institut f\"ur Kernphysik, Forschungszentrum J\"ulich, J\"ulich,
Germany \\
$^2$ Centro de F\'isica Nuclear, Universidade de Lisboa, 1649-003 Lisboa,
Portugal \\
$^3$ Department of Physics and Astronomy, University of South Carolina,
Columbia, SC 29208, USA \\
$^4$ Department of Physics and Nuclear Theory Center,
Indiana University, Bloomington, IN 47405, USA \\
$^5$ Department of Physics, University of Washington, Seattle,
WA 98195-1560, USA \\
$^6$ Department of Physical Sciences, University of Helsinki,
Helsinki, Finland \\
$^7$ Department of Physics, University of Arizona, Tucson,
AZ 85721, USA}

\begin{abstract}
We use realistic two- and three-nucleon
interactions in a hybrid chiral-perturbation-theory calculation
of the charge-symmetry-breaking reaction $dd\to\alpha\pi^0$ to 
show that
a cross section of the experimentally measured size can be obtained using
LO and NNLO pion-production operators.
  This  
result  supports
the validity of our power counting scheme 
and demonstrates the necessity of using  an accurate treatment
of ISI and FSI.
\end{abstract}
\end{frontmatter}

{\bf 1.} For most  purposes, hadronic isospin states can be considered as
charge symmetric, \IE, invariant under a rotation by $180^\circ$ around the
2-axis in isospin space.
Charge symmetry (CS) is a subset of the general isospin symmetry, charge
independence (CI), which requires invariance under any
rotation in isospin space.
In quantum chromodynamics (QCD), CS implies  that  dynamics are invariant 
under the exchange of the up and down quarks~\cite{Miller:1990iz}.
However, since the up and down quarks do have different masses
($m_u\neq m_d$)~\cite{weincsb,mumd}, 
the QCD Lagrangian is not charge symmetric.
This symmetry violation is called charge symmetry breaking (CSB).
The different electromagnetic interactions of the up and down quarks
break CI.  Observing the effects of CSB interactions therefore provides
a probe of $m_u$ and $m_d$.

\vspace{0.1cm}
Two exciting recent observations of CSB in experiments
involving the production of neutral pions stimulate our attention.
Many years of effort  led to the observation of CSB in $np\to d\pi^0$ at
TRIUMF.
The CSB forward-backward
asymmetry of the differential cross section was found to be
$A_{\rm fb}=[17.2\pm8({\rm stat})\pm5.5({\rm sys})]\times 10^{-4}$
~\cite{Opper:2003sb}.
In addition, the final experiment at the IUCF Cooler ring reported
a very convincing $dd\to\alpha\pi^0$ signal near threshold
($\sigma=12.7\pm2.2$~pb at $T_d=228.5$~MeV and $15.1\pm3.1$~pb at 231.8~MeV)
\cite{Stephenson:2003dv}.
The $dd\to\alpha\pi^0$ reaction 
violates CS since the deuterons and the $\alpha$-particle
are self-conjugate under the CS operator, with a positive
eigenvalue, while the neutral pion wave function changes sign.

\vspace{0.1cm}
The study of CSB $\pi^0$  production reactions
presents an exciting new opportunity to  learn about the influence of 
quark masses in nuclear physics, and to use  effective
field theory (EFT) to improve 
 our understanding of 
how QCD works~\cite{mosreview}.
This is because chiral symmetry of QCD determines the form of pionic 
interactions.
 Electromagnetic CSB  is typically of the same
order of magnitude as the strong one,  and also can be handled using EFT.

The EFT for the Standard Model at momenta comparable to the pion mass, $Q\sim
m_\pi$, is chiral perturbation theory ($\chi$PT)~\cite{ulfreview}.
This EFT has been extended 
\cite{cpt0,vKNM,hanhartreview,Gardestig:2005sn,Gardestig,Lensky:2005jc}
to momenta relevant to
pion production, $Q\sim \sqrt{m_\pi M}$ with $M$ the nucleon mass.
(For a review and further references, see Ref. \cite{hanhartreview}.)

\vspace{0.1cm} 
EFT 
with the operators of Ref. \cite{vkiv}
was used to correctly predict the sign of
the forward-backward asymmetry in $np\to d\pi^0$ \cite{vKNM}.
For the  $dd\to\alpha\pi^0$ reaction, we 
surveyed various mechanisms using  initial-state plane-wave functions 
and simplified final-state wave functions \cite{Gardestig}.
In this simplified model, we found that the formally leading-order (LO)
production mechanism is suppressed through symmetries in the 
wave functions and studied other mechanisms.
The contributions from next-to-next-to-leading-order
(NNLO) diagrams are too small to account
for the observed cross section --- a cross section of only $0.9$ pb was found.
We  also included 
short-range pion emission, which contributes at N$^4$LO
through contact vertices whose strengths are a priori unknown.
We used resonance saturation,
by means of 
CSB effects in Z--diagrams, as motivated by a
successful phenomenological  model \cite{lowe} 
of the  
charge-symmetry-conserving (CSC) reaction $pp\to pp\pi^0$. For the 
simplified wave functions, we then found a  
cross section of the observed order of magnitude.

\vspace{0.1cm}
Our aim here is to take advantage of recent significant
advances in  four-body theory \cite{nogga,Fonseca99}
that allow us to
include the effects of  deuteron-deuteron interactions in the initial state,
and to use bound-state wave functions with realistic two- and three-nucleon
interactions. 
The calculations presented here are hybrid: the pion-production operators
are constructed using EFT, but the nuclear 
interactions used to obtain the wave functions are not.
No calculation of this kind can be considered to be completely well-founded
unless the operators and wave functions are
constructed from the same convergent EFT. However, the interactions
do include 
one-pion exchange, so  their long-range
behavior is founded in EFT. Moreover, we   employ 
several potentials
to  
gauge the sensitivity of the various
production operators  to 
 the  shorter-ranged parts of the interaction.

\vspace{0.1cm}
The present  study does not include all 
diagrams appearing
at NNLO.  A complete analysis 
demands the inclusion of  loop diagrams.
 Their evaluation requires a careful treatment of divergences
as pointed out in Ref. \cite{Gardestig:2005sn}, and  understood only recently
\cite{Lensky:2005jc}. For technical reasons, 
photon exchange is so far only considered in
the final-state wave function.
In this letter, we concentrate on the important effects of 
initial- and final-state interactions (ISI and FSI), and 
anticipate  that 
the use of the present
incomplete set of operators should be sufficient to get order-of-magnitude estimates 
and to demonstrate the technique.

\begin{figure}[t!] 
\vspace{3.cm}
\begin{center}
\includegraphics{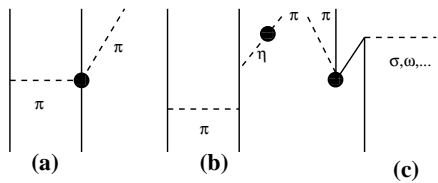}
\caption{Diagrams of  $np\to d\pi^0$; the solid circle  indicates CSB.}
\label{ddlettdiag}
\end{center}
\end{figure}

\vspace{0.1cm}
{\bf 2.}
We summarize 
the power counting  of the previous study \cite{Gardestig}, which contains
explicit expressions for the operators.
At LO,
there is only one contribution,
represented by Fig.~\ref{ddlettdiag}a:
pion rescattering in which the
CSB occurs through the seagull pion-nucleon terms
linked to the nucleon-mass splitting.
This contribution stems from the chiral transformation properties of the
quark operators that generate CSB, which are two:
{\it i)} the up-down mass difference, which breaks chiral symmetry as a
component of a chiral four-vector;
{\it ii)} electromagnetic quark interactions, which break chiral symmetry
as components of a chiral anti-symmetric rank-two tensor. 
In lowest order, there exist two seagull operators involving a
nucleon interacting with two pions, one of which is neutral.
Their strengths are determined by the quark-mass and electromagnetic
contributions to the nucleon mass splitting,
$\delta m_N$ and $\bar{\delta}m_N$, respectively.
$\delta m_N$ is proportional to $\varepsilon (m_u+m_d)$, where
$\varepsilon \equiv (m_u-m_d)/(m_u+m_d) \approx 1/3$,
while $\bar{\delta}m_N$ is the fine-structure constant $\alpha$
times a typical hadronic mass.
The only existing constraint on these two terms 
is
$\delta m_N+\bar{\delta}m_N
=M_n-M_p=1.29~{\rm MeV}$ and the model dependent estimate 
$\bar{\delta}m_N=-(0.76\pm0.3) \;{\rm  MeV}$ 
based on the Cottingham sum rule \cite{gl}. 
Verifying the theory requires that the two terms
be constrained  independently.
The most natural reaction to study is $\pi N$ scattering.
Ref.~\cite{weincsb} predicted a significant difference between the 
$\pi^0 p$ and $\pi^0n$ scattering lengths
that is not presently observable,
as discussed in Ref. \cite{nadjaIV}. 
Effects of these terms in
the nuclear potential \cite{CSBNN} are relatively small or suffer from other 
unknowns as in $\pi d$ scattering \cite{urr}. 
This leaves the investigation of CSB in the two reactions,
$np\to d\pi^0$ and $dd\to \alpha\pi^0$, as  very promising
possibilities.
For definiteness,
in this paper we use the central value of the estimate from the 
Cottingham sum rule.
The leading diagram, Fig.~\ref{ddlettdiag}a,
is $O[\varepsilon m_\pi^2/(f_\pi^3 M Q)]$,
where $f_\pi =92.4$ MeV
denotes the pion decay constant and $Q \approx \sqrt{m_\pi  M}$ a typical 
momentum.

We refer to this contribution as ``pion exchange''. For completeness, we distinguish the parts 
proportional to  $\delta m_N$ 
and $\bar {\delta} m_N$ 
and denote the contributions  
by ${\mathcal M }_{PE}={\mathcal M }_{PE, \delta m_N}+{\mathcal M }_{PE, \bar {\delta} m_N}$.

\vspace{0.1cm}
There is no NLO 
contribution.  
At NNLO, suppressed by $O(m_\pi/M)$,
there exists a recoil correction
 of the LO term (labeled 
 ${\mathcal M}_{rec} = {\mathcal M}_{rec,\delta m_N}+{\mathcal M}_{rec,\bar{\delta}m_N}$).
Its strength is also determined by  $\delta m_N$ and $\bar{\delta}m_N$.
Therefore, its contribution
allows us to estimate the size of NNLO contributions.
  The recoil correction to the $\pi NN$ vertex is linear in the energy of the
  virtual pion. Such operators were studied in
  Ref. \cite{toy} for the reactions $NN\to NN\pi$,
 and we  use the prescription provided there and applied
  in Ref. \cite{Gardestig}. Demonstrating  the validity of this
 recipe  for a four-body
  environment deserves   further study.

\vspace{0.1cm}
At the same order new parameters appear.
In particular, a term arises in which
a one-body CSB operator ($\propto \beta_1+\bar{\beta}_3)$
is sandwiched between initial- and final-state wave functions, as illustrated
in, {\it e.g.}, Fig.~\ref{ddlettdiag}b. 
We refer to this  as the one-body term (${\mathcal M}_{1b}$).
The terms $\beta_1 = O(\epsilon m_\pi^2/M^2)$ and
$\bar\beta_3 = O(\alpha/\pi)$ arise from, respectively,
the quark-mass-difference and
electromagnetic contributions to the isospin-violating pion-nucleon coupling.
Neither $\beta_1$ nor $\bar\beta_3$ can be extracted from experiment
yet. To allow us to provide numerical results we
 estimate these terms by modeling~\cite{vKFG} $\beta_1$ 
by $\pi$-$\eta$ mixing, see Fig.~\ref{ddlettdiag}b,
\begin{equation}
\beta_1=\bar{g}_\eta\langle\pi^0|H|\eta\rangle/m_\eta^2, \label{beta1}
\end{equation}
where $\langle\pi^0|H|\eta\rangle=-4200$~MeV$^2$
is the $\pi$-$\eta$--mixing matrix
element~\cite{mesmix}, and
$\bar{g}_\eta=g_{\eta NN} f_\pi/M$
the $\eta$-nucleon coupling constant.
An early analysis
\cite{ODu83} using one-boson-exchange potentials in $NN$ scattering gave 
$g_{\eta NN}^2/4\pi=3.86$ (used in Ref. \cite{vKNM}),
but the data show  little sensitivity to $\eta$ exchange and
high-accuracy fits
can be achieved \cite{Machleidt:2000ge} using  $g_{\eta NN}^2/4\pi=0$.
Indeed, the
possibility  of a vanishing coupling constant had been  raised earlier.
The detailed analysis of $NN$ total cross sections and
$p\bar{p}$ data using dispersion relations \cite{Grein:1979nw}
found  that $g_{\eta NN}^2/4\pi=0$.
This is consistent with extractions from the nucleon pole in the amplitude
$\pi N\to\eta N$ that give \cite{Deans:69} $0.5>g_{\eta NN}^2/4\pi\ge0$.
Photoproduction reactions on a nucleon \cite{getasmall} (see their Fig.~2)
yield
the  small value   $g_{\eta NN}^2/4\pi=0.1$.  To be consistent
with our earlier study \cite{Gardestig}, we use $g_{\eta NN}^2/4\pi = 0.51$
for the results shown below, but also examine the effects of using 
$g_{\eta NN}^2/4\pi = 0.10$.
Both values are roughly consistent with  the size expected using  
power counting arguments \cite{vKFG}.
We assume the sign predicted by $SU(3)$ symmetry, as in Ref. \cite{Gardestig}.

\vspace{0.1cm} 
The effects of electromagnetic 
interactions as well as strong CSB were included in computing the
 $\alpha$-particle wave functions, where the former effect is  dominant. 
 These  interactions  generate a small
isospin $T=1$ component of the wave function that  enables
a non-zero contribution of CSC production operators. To estimate the effects of
the admixtures, we calculate the production matrix element using the CSC
counterpart of diagram Fig.~\ref{ddlettdiag}b 
(referred to as ${\mathcal M}_{WF}$). 

\vspace{0.1cm}   
A number of other CSB mechanisms enter at N$^3$LO or higher,
including additional loop diagrams and short-range interactions.
The lowest order where four-nucleon contact interactions start to contribute
is N$^4$LO, that is, $O(m_\pi/M)$ below NNLO. 
To estimate their strength,
Ref.~\cite{Gardestig}
evaluated certain tree-level contributions as indicated by
Fig.~\ref{ddlettdiag}c, which
represents the exchange of heavy mesons
($\sigma$, $\omega$, $\rho$) via a Z-graph mechanism,
with $\pi$-$\eta$ mixing generating CSB at pion emission
(${\mathcal M}_\sigma$, ${\mathcal M}_\omega$  and ${\mathcal M}_\rho$).
Another Z-graph (labeled as ${\mathcal M}_{\rho\omega}$) 
arises in which the CSB occurs  in the heavy-meson exchange via 
 $\rho$-$\omega$ mixing along with strong  pion emission at the vertex.
The Z-graphs are believed to be important because their inclusion leads to a
quantitative description of the total cross section for the reaction $pp\to
pp\pi^0$ near threshold 
\cite{lowe}. Our  present results use the coupling constants  and parameters of
Ref.~\cite{Gardestig}, see their Table I. However, in the future it will 
be necessary to reassess the procedure in light of recent developments 
concerning the treatment of divergences in EFT loop 
diagrams \cite{Lensky:2005jc}. 

\vspace{.2cm} {\bf 3.} The various mechanisms generate pion-production kernels
that are sandwiched between final- and initial-state wave functions  
to provide a
transition matrix element ${\mathcal M}$.  
We restrict our analysis to $T_d=228.5$ MeV, as the
effects of a small change in energy are captured mainly by the change in
the phase-space factor.  
The cross section is related to the matrix elements by
\be
\sigma=4.303 \;{\rm pb} \  \left\vert {\mathcal M}\; 10^4\ {\rm fm}^{2}\right 
\vert^2.
\label{cross}
\ee

\vspace{0.1cm}
We present our new results in stages.
First we introduce realistic bound-state wave functions,
while continuing to use the plane-wave approximation (PWA).
The techniques to solve the four-body problem have been presented
by Nogga {\it et al.} \cite{nogga}.
To be specific, we present results using
both the 
AV18 \cite{Wiringa:1994wb} and CD-Bonn~2000  \cite{Machleidt:2000ge}
two-nucleon potentials  combined with a properly adjusted 
Tucson-Melbourne (TM99) \cite{tm99} three-nucleon force. 
The combination guarantees
that the $\alpha$-particle binding energy is 
reproduced with high accuracy. Additional calculations
using the Urbana-IX \cite{Pudliner:1997ck} three-nucleon potential resulted in
essentially identical results and will be presented elsewhere \cite{ustbp}.

\vspace{0.1cm}
Table~\ref{tablecal} summarizes our results
for the transition amplitudes 
that add to ${\mathcal M}$, 
labeled according to the
various mechanisms described above.
The one-body term is predicted rather model independently. Using  these
matrix elements for the one-body operator leads to a
 cross section of $10-13$~pb, which is accidentally in good agreement
with the experiment. Compared to our toy-model calculation \cite{Gardestig}, 
we find
an increase of the cross section by a factor of 10, showing that the 
high-momentum tail of the wave function is important, 
especially for the one-body term.
Using the smaller, but also realistic, coupling $g_{\eta NN}^2/4\pi = 0.10$ would reduce
the resulting cross section by a factor of 5. 
The one-body term is formally subleading. 
However, the toy-model calculation showed that the pion-exchange 
term is suppressed
due to the symmetry of the $\alpha$-particle wave function.
This result persists
for the realistic $\alpha$-particle wave functions: the amplitude does not
vanish exactly,
but remains smaller than the one-body term. 
This term is quite sensitive to the chosen nuclear interaction, pointing
to sensitivity to the short-range part of the potential
and to the small
components of the $\alpha$-particle wave function.

\begin{table}[tb]
\caption{Complex
 $dd\to\alpha\pi^0$ amplitudes at $T_d=228.5$ MeV  in   units of
$10^{-4}$ fm$^{-2}$. PWA denotes the plane-wave approximation. 
ISI results also include the initial-state interaction.}
\begin{tabular}{|l|cc|cc|} \hline\hline
& \multicolumn{2}{c|}{PWA} & \multicolumn{2}{c|}{ISI} \cr
 & CDB+TM99 & AV18+TM99  &  \quad CDB+TM99 \qquad & \quad AV18+TM99 \qquad    \cr      
\hline        
${\mathcal M}_{PE,\delta m_N}$              
                                        & 0.35    & $-0.07$    &  $-1.51$ + i  1.87  &  $-0.76$ + i  0.74  \cr
${\mathcal M}_{PE,\bar{\delta} m_N}$     
                                       & 0.06    & $-0.01$    &  $-0.28$ + i  0.35  &  $-0.14$ + i  0.14  \cr
${\mathcal M}_{PE}$              & 0.41    & $-0.08$    &  $-1.79$ + i  2.22  &  $-0.90$ + i  0.88  \cr
\hline
${\mathcal M}_{rec, \delta m_N}$                
                                         & 0.41    & 0.34      &  $-0.81$ + i 0.74   &  $-0.63$ + i 0.59  \cr
${\mathcal M}_{rec, \bar{\delta} m_N}$                
                                         & 0.08    & 0.06      &  $-0.15$ + i 0.14   &  $-0.12$ + i 0.11  \cr
${\mathcal M}_{rec}$                & 0.49    & 0.40      &  $-0.96$ + i 0.88   &  $-0.75$ + i 0.70  \cr
\hline                 
${\mathcal M}_{1b}$                 & 1.76     & 1.60     &  $-2.51$ + i 1.84   &  $-1.94$ + i 1.60  \cr
\hline                 
${\mathcal M}_\sigma$           & 0.46    & 0.31     &  $-0.56$ + i 0.64   &  $-0.32$ + i 0.42  \cr
${\mathcal M}_\omega$            & 0.51    & 0.38      & $-0.53$ + i 0.44   &  $-0.35$ + i 0.34   \cr
${\mathcal M}_\rho$                  & 0.24    & 0.15     & $-0.33$ + i 0.34    &  $-0.18$ + i 0.19   \cr
${\mathcal M}_{\rho\omega}$   & 1.17    & 0.87      & $-1.32$ + i 1.51    &  $-0.84$ + i 1.07  \cr
\hline                     
${\mathcal M}_{WF}$                & $-0.15$   & $-0.14$    & +0.51 $-$ i 0.13    &  +0.41 $-$ i 0.14  \cr
\hline\hline
\end{tabular}
\label{tablecal}
\end{table}

\vspace{0.1cm}
Since the LO term is suppressed and the one-body term
is not well constrained by resonance saturation, it is interesting to look
at the pion-recoil term. Its parameters are better determined than
$\beta_1$, since they are related to the nucleon mass difference and the
Cottingham sum rule. Our calculation may 
therefore give a trustful estimate
of the size of the NNLO contributions. The results are a rather 
model-independent amplitude of approximately  $1/3$ the size of the 
one-body term and are  
in line with the power counting.

\vspace{0.1cm}
In contrast,
we find that all the Z-graphs give unexpectedly large
contributions, especially the $\rho$-$\omega$ exchange operator. Also, the
contributions add constructively, so that their sum tends to overwhelm
the one-body term. This model of  resonance saturation thus gives 
 results in vast disagreement with the power counting. 

\vspace{0.1cm}
CSB effects on the final-state wave functions (${\mathcal M}_{WF}$) are smaller
than pion recoil 
and insensitive to the chosen nuclear interaction, 
indicating
that these terms are well constrained using phenomenological interactions.

\vspace{0.1cm}
{\bf 4.} The next step is  to present  the effects of
  including the ISI.
The correct treatment of this involves the solution of the four-body
scattering problem at center-of-mass energies greater than 200 MeV.
In spite of the tremendous progress achieved in recent years on the
solution of the four-nucleon problem \cite{Ciesielski99,Fonseca99},
advances in obtaining exact solutions are limited to 
energies below the four-particle breakup threshold. To understand our
pion-production reaction it is necessary to  go beyond the
distortions obtained through an effective optical-model potential
fitted to the elastic ${\vec d}d$ scattering data.
 This is because important pion production occurs in which the
deuterons interact, break up, and then emit a pion. At very high
energies the use of Glauber approximation is justified, but in the
threshold energy regime for pion production the wave length
associated with the relative $d+d$ on-shell momentum is close to the
size of the deuteron. Therefore we obtain an approximate solution of
the Yakubovsky \cite{Yakubovsky67} equation for the four-nucleon
scattering wave function that is made up of two terms: the first
involves the bound-state wave functions of the two deuterons times a
plane wave  describing the relative motion between them; the second
requires the breakup of one of the deuterons followed by the
three-body scattering of the $N+d$ system into the three-particle
continuum in the presence of the remainder spectator nucleon.

\vspace{0.1cm} 
Such an approximation is based on the lowest-order terms
in the Neumann series expansion of the four-particle Yakubovsky
equation, leading to the following expression for the scattering
wave function,
\begin{eqnarray} \label{eq:Psi3}
|\Psi^{\rho_{_0}}\rangle  
\simeq|\phi^{\rho_{_0}} \rangle  
  +  \sum_j \sum_{i\rho} G_{_0} \; t_i\;
G_{_0} \; U^\rho_{ij} \; \bar\delta_{\rho\rho_{_0}}|
\phi^{\rho_{_0}}_j \rangle\ ,
\end{eqnarray}
 where $\rho$ denotes one of the
seven two-body partitions, four of $(3) + 1$ type and three of $(2)
+ (2)$ type, and $i$ is a pair interaction that is internal to
$\rho$; $j$ is both internal to $\rho$ and $\rho_{_0}$. 
The initial-state wave function component $|\phi^{\rho_0}_j\rangle$ carries the
appropriate bound-state wave function components of the target and
projectile times a relative plane wave between their respective
center of mass. As usual, $\rho_{_0}$ specifies the two-body
entrance channel, $\bar \delta_{\rho\rho_{_0}} = 1 -
\delta_{\rho\rho_{_0}}$, $G_{_0}$ is the four-free-particle Green's
function and $t_i$ the $t$-matrix for pair $i$. If $\rho_{_0}$
corresponds to a $2+2$ initial state, then $\rho$ can only be a
$(3)+1$ two-body partition and $U^\rho_{ij}$ is  the solution of the
three-body Alt, Grassberger and Sandhas (AGS) \cite{Alt70} equation
for the three-particles that make up subsystem $\rho$.
The first term in Eq.~(\ref{eq:Psi3}) corresponds to the
initial-state wave function; the second term requires the breakup of
one of the bound pairs followed by the scattering of either one of
the particles from the remaining bound pair, leading to four free
particles in the continuum. The particle-pair scattering into the
continuum takes place in the presence of the fourth one, and
therefore, by energy conservation, the total energy available is the
total four-body center-of-mass energy minus the relative kinetic
energy of the fourth particle relative to the center of mass of the
other three. Thus the four-body scattering wave function we
construct contains all orders in the pair interaction but also
three-particle correlations in first-order perturbation at all
possible energies that are consistent with four-particle energy
conservation.

 \vspace{0.1cm}
For four identical nucleons Eq.~(\ref{eq:Psi3}) may be written as
\begin{eqnarray} \label{eq:Psidd}
|\Psi^+_{dd}\rangle  =  |\phi_{dd}\rangle
                  + \frac{1}{\sqrt{12}} \;(1 + P -P_{34} P + \tilde{P} )
                  (1 - P_{34}) |\psi^{(12,3)4}_{dd}\rangle ,
\end{eqnarray}
where $P = P_{12} P_{23} + P_{13} P_{23}$ and $\tilde{P} = P_{13}
P_{24}$ are permutation operators whose appropriate combination
generates the 6 (12) components of the Yakubovsky wave function that
are of $2+2$ ($1+3$) type. The symmetrized $d+d$ initial-state wave
function
 \cite{Gardestig} $|\phi_{dd}\rangle$ is given by
\begin{eqnarray} \label{eq:Phidd}
|\phi_{dd}\rangle = \frac{1}{\sqrt6}\;
           (1 +  P - P_{34} P + \tilde{P})
           \; \xi_d (12)\; \xi_d (34) \; {\rm Exp} (12-34) ,
\end{eqnarray}
where $\xi_d(ij)$ is the deuteron wave function for the pair $(ij)$,
and Exp$(12-34)$ represents the relative plane wave between the two
deuteron pairs. The second term in Eq.~(\ref{eq:Psidd})  mandates
the use of a  specific choice of wave-function component for
nucleons 1, 2, 3 and 4,
\begin{eqnarray} \label{eq:Psidd34}
|\Psi^{(12,3)4}_{dd}\rangle   = G_0 \, \langle 12,3|\; U_0(z)| (12)3
\rangle \,
                               \xi_d(34) ,
\end{eqnarray}
where connectivity increases from left to right and
$z=E-\frac{4}{3}\;k^2+i0$, $k$ being the relative momentum between
nucleon 4 and the center of mass of (123). The breakup operator $U_0
= t G_0 U$ and the corresponding matrix element $\langle
12,3|\;U_0(z)| (12)3 \rangle $ represents the scattering of nucleon
3 from the bound state of (12) leading to three free nucleons in the
continuum where nucleons 1 and 2 are last to interact through their
respective $t$-matrix $t$. The operator $U$ is the AGS three-body
scattering operator
 that satisfies the integral equation
\begin{eqnarray} \label{eq:U}
U = P G^{-1}_0 + P t G_0 U,
\end{eqnarray}
from which one calculates $Nd$ elastic scattering amplitudes.
The permutation operator $P$ is the same  as used in
Eq.~(\ref{eq:Psidd}) and corresponds to the sum of the two cyclic
permutations of particles 1, 2 and 3.
 We extract from Eq.~(\ref{eq:Psidd}) the $^3$P$_0$ partial
wave in the entrance channel to compute the threshold cross section
for $dd \to \alpha \pi^0$.

\vspace{0.1cm}
Details regarding the numerical solution of the scattering problem
and evaluation of the pion-production matrix-element
will be presented elsewhere \cite{ustbp}. 
Here, in Table \ref{tablecal}, we simply present the
results for 
the transition amplitudes, 
which acquire imaginary parts due to the
presence of the initial-state interaction.
Generally, we observe a significant enhancement of all contributions to the
amplitude.

\vspace{0.1cm}
It is immediately apparent that the pion-exchange term,
which is supposed to be LO, is now of the size
of the NNLO terms considered here, 
namely the 
pion-recoil 
and one-body operators.
It still shows a sizable model dependence, which 
could visibly influence our final result for 
the cross section.
A more consistent treatment of
nuclear interactions and production operators
will be necessary in future.

\vspace{0.1cm}
The 
pion-recoil and one-body terms remain relatively model
independent. Both, therefore, can serve as an order-of-magnitude
estimate of the cross section. Our results for these matrix elements
correspond to cross sections between 4.5 and 42 pb. Again, the one-body
contribution is larger than the pion-recoil term. Both would come
close to each other for the smaller choices of the strength
of the one-body term. 
Our explicit calculation shows that the NNLO contribution provides
a strength consistent with the experiment. We stress that the strong
enhancement due to initial-state interactions and higher-momentum
components of the $\alpha$-particle wave functions are necessary to
find NNLO contributions of the required size. 

\vspace{0.1cm}
The ISI
also enhances short-range contributions
from Z-graphs, but by far-smaller amounts.
We still find relative contributions much larger than expected from
the power counting and, again, all the contributions add up constructively.
One possible explanation would be that there is simply no convergent EFT for
the reaction $dd\to \alpha \pi^0$ and the Z-graphs would need to be included
as done in this work. However, if this was true the value of the
empirical cross section would be the result of subtle cancellations amongst
various terms from very different origins, a very unlikely coincidence.   
The  more likely interpretation is that the power counting
works, but the Z-graphs simply
provide the wrong model to estimate the four-nucleon operators. If this is the
case we may even drop them all together from our investigations, since they
are of 
high order. 
      
\vspace{0.1cm} 
{\bf 5.} This paper  extends our earlier study of the
reaction $dd\to\alpha\pi^0$ by using  realistic wave functions for the 
four-nucleon ISI and FSI.
We also provide numerical estimates for some diagrams
that lead to a cross section of the right order of magnitude supporting 
the power counting given the suppression of the LO. 
In addition, the
present results
allow us to identify a few issues that deserve further study
(in addition to
the inclusion of all diagrams up to NNLO).

\begin{itemize}
\item Given the dramatic influence of initial-state interactions,
 it is of paramount importance that new experimental
  constraints be obtained for the  deuteron-deuteron interactions
  in the energy region close to the pion-production threshold.  Besides data
  on elastic $dd$ scattering, also data on other pion-production reactions with
  the same initial $^3$P$_0$ state are needed. 
  The most obvious examples are the CSC
  reactions $dd\to ^3 {\rm H/He} \, N\pi$
    recently measured at COSY 
\cite{anke},
which are an interesting alternative
  to elastic $dd$ scattering because only  a few partial waves  contribute
  in the entrance channel.
 
\item Another  important issue is to better understand the role of
  the Z-graphs
  used to estimate the size of four-nucleon operators. As shown
  above, their size is much larger than indicated by their N$^4$LO 
  power-counting order. 
  Thus, it is necessary to reassess the procedure in light of
  recent developments in EFT, especially concerning the treatment of
  divergences in loop diagrams \cite{Lensky:2005jc}.  
   Given the large number of experimental data, especially for
   $pp\to pp\pi^0$, much insight can be obtained. 
 \item Finally, we found some dependence on the chosen 
 nuclear interaction model.
 This is probably related to the presumed inconsistency between the
 chosen 
 nuclear interaction and the production operators. 
 Such an inconsistency can only be resolved by applying nuclear interactions
 based on chiral perturbation theory \cite{ulfreview,NNChPTold} that need 
 to be extended to higher cutoffs as outlined in Ref. \cite{Nogga:2005hy}.
\end{itemize}

\vspace{0.1cm} 
The central  result of this letter is that the inclusion of ISI and FSI
enhances the contribution of NNLO diagrams of the production operator
enough to able to account for the measured cross section. Together
with the insight that the LO is suppressed, this supports the EFT approach
to this reaction. But it also demonstrates that a careful treatment of the
nuclear effects is required for a final analysis.

\vspace{0.1cm} 
In view of a planned  
measurement of the reaction $\vec dd\to \alpha\pi^0$ at higher
energies at COSY \cite{wasa} ---where $p$ waves will be relevant---
and 
of the experimental determination 
of $A_{fb}$ \cite{Opper:2003sb},
we will have an increased database on cross sections, which will allow
us to disentangle the various  contributions. 
Although much remains to be done before any precise statements about
the values of the
parameters $\delta m_N$, $\bar{\delta} m_N$ can be made, we
are now at threshold of understanding how the
light-quark masses makes a difference in nuclear physics.

\section*{Acknowledgments}

We thank Andy Bacher, Allena Opper, and Ed Stephenson
for useful discussions and encouragement, without which this work would
not have been completed.
This research was partially funded by
FCT grant POCTI/37280/FNU/2001 (ACF),
NSF grant PHY-0457014 (AG),
DOE grants DE-FG02-97ER41014 (GAM) and DE-FG02-04ER41338 (UvK),
and a Sloan Research Fellowship (UvK).

\bibliographystyle{unsrt}

\end{document}